\documentclass{article}
\usepackage{ijcai18}

\usepackage{times}
\usepackage{xcolor}
\usepackage{graphicx}
\usepackage{float}
\usepackage{amsmath}
\usepackage{enumitem}
\usepackage{soul}
\usepackage{array}
\usepackage[utf8]{inputenc}
\usepackage[small]{caption}
\usepackage{booktabs}
\usepackage{hyperref}
\usepackage{fancyhdr}
\title{Polyphonic Music Composition with LSTM Neural Networks and Reinforcement Learning}

\author{
Harish Kumar$^1$, 
Balaraman Ravindran$^2$, 
\\ 
$^1$ Texas A\& M University \\
$^2$ Indian Institute of Technology, Madras \\
harishk1908@gmail.com,
ravi@cse.iitm.ac.in
}

\begin{document}

\maketitle

\setlength{\abovedisplayskip}{2pt}
\setlength{\belowdisplayskip}{2pt}
\setlength{\abovedisplayshortskip}{2pt}
\setlength{\belowdisplayshortskip}{2pt}

\begin{abstract}
  In the domain of algorithmic music composition, machine learning-driven systems eliminate the need for carefully hand-crafting rules for composition. In particular, the capability of recurrent neural networks to learn complex temporal patterns lends itself well to the musical domain. Promising results have been observed across a number of recent attempts at music composition using deep RNNs. These approaches generally aim at first training neural networks to reproduce subsequences drawn from existing songs. Subsequently, they are used to compose music either at the audio sample-level or at the note-level. We designed a representation that divides polyphonic music into a small number of monophonic streams. This representation greatly reduces the complexity of the problem and eliminates an exponential number of probably poor compositions. On top of our LSTM neural network that learnt musical sequences in this representation, we built an RL agent that learnt to find combinations of songs whose joint dominance produced pleasant compositions. We present \textbf{Amadeus}, an algorithmic music composition system that composes music that consists of intricate melodies, basic chords, and even occasional contrapuntal sequences. 
\end{abstract}

\section{Introduction}

Computer-driven music composition has been approached through various pathways (\cite{fernandez2013ai}): through knowledge-based systems, evolutionary algorithms, Markov Processes, and in the last 25 years, Artificial Neural Networks(ANNs). The presence of a strong mathematical framework behind music has explicitly and implicitly aided researchers in these approaches. ANNs' propensity to successfully learn complex patterns from raw data has made them especially suited to the problem of algorithmic music composition. Several research endeavours have been undertaken on the subject of training ANNs to compose original music that appeals to human sensibilities. These works have resulted in computer systems that can fashion musical sequences that are occasionally quite difficult to discern from human-composed music. \par
Despite these advances, the application of Artificial Intelligence in music composition is a nascent domain that has great potential for extensive scientific research. In this paper, our contributions to this evolving area are as follows: We have built a novel representation for polyphonic music that has a much lower sparsity when compared to other existing representations. Attempts at using reinforcement learning towards getting preferred compositions from already trained neural networks have been very few - we take an approach that is quite different from these attempts and present the results of our work where we observe significant improvements when using RL. We also present, to the best of our knowledge, the first instance of a method to intelligently explore the powerful {\em plan} space first introduced by \cite{todd1989connectionist}.
\par

\section{Conventions and Definitions}

We use the following set of conventions and definitions throughout this paper:

\begin{itemize}
\setlength\itemsep{-0.1em}
\item A {\em note} is equivalent to playing a pitch for a definite duration.
\item A {\em note-set} is a set of pitches that begin simultaneously, but last for independently defined durations.
\item Unless specified otherwise, the term \textbf{subsequence} means a contiguous subsequence.
\item Musical sequences are visually presented in the piano roll visualization\footnote{While this has its own limitations, it greatly aids in appreciating the general flow of the melody, harmony, rhythm, etc.}. The vertical axis represents note pitch while the horizontal axis represents the flow of time. 
\end{itemize}

\section{Related Work}

\cite{todd1989connectionist} trains a neural network to first learn sequences drawn from existing music. Input configurations called {\em {\em plans}} are used to encode the identity of sequence that the neural network is presently learning. During the composition phase, interpolated {\em plans} are used to generate new melodies. \cite{mozer1991connectionist} use a Recurrent Neural Network along with a multidimensional pitch representation based on psychological studies from \cite{shepard1982geometrical}. Both these approaches are examples of using an ANN to generate a melody sequentially, note-by-note. \par
There has also been research (\cite{fernandez2013ai}) into harmonizing existing melodies with \cite{shibata1991harmony} and \cite{melo1998connectionist} being some instances. \cite{liang2016bachbot} and \cite{hadjeres2017deepbach} demonstrate a neural network harmonizing three out of the {\em soprano}, {\em alto}, {\em tenor} and {\em bass} parts of a Bach chorale when one of them is fixed. \par
Numerous attempts at sequential music composition with neural networks have used the Long Short-Term Memory units introduced in \cite{hochreiter1997long}, starting with\cite{eck2002first}. \cite{colombo2016algorithmic} in contrast, apply Gated Recurrent Units (GRU) for monophonic composition. \par
There are a few instances(\cite{franklin2001multi} and \cite{jaques122016tuning}) of reinforcement learning being used over trained neural networks to impose additional conditions on the composed music. Another previously used non-reinforcement strategy for imposing external conditions on a separately trained neural network is to use sampling grammars as seen in \cite{sun2016composing}.

\section{LSTM Neural Networks}

The temporal capacity of a reccurent neural network allows it to learn patterns that are spread over time, and this has permitted RNNs to be used in signal processing, natural language processing and music generation among other areas. However, simple recurrent neural networks suffer from the vanishing and exploding gradients problems demonstrated in \cite{bengio1994learning}. These problems cause the gradients corresponding to earlier inputs to either vanish or blow-up depending upon their value, and thus these networks have difficulty in learning long-range temporal dependencies. This is a particular deficiency for learning patterns in music as temporal dependencies in musical sequences may be several notes apart. \par
Long-Short Term Memory units (\cite{hochreiter1997long} and \cite{Gers99learningto}) step around this problem by using a constant error carousel that trap the error within a cell. Two gating neurons regulate the flow of information into a cell and out of it respectively, while one neuron learns whether to forget the value inside the cell. \par
Specific details about the input, output and internal state relations followed in LSTM units can be found in \cite{hochreiter1997long} and \cite{Gers99learningto}. 

\section{Input Representation}

\subsection{Pitch and Duration Encodings}
We first transpose the musical sequences to the same key, in accordance with positive results observed in \cite{mozer1991connectionist}, \cite{franklin2004recurrent}, \cite{sun2016composing}, \cite{boulanger2012modeling} and \cite{liang2016bachbot}.\par
In contrast with the approaches in \cite{mozer1991connectionist} and \cite{franklin2004recurrent} where encodings based on musical information were used to represent pitches, chords and note durations, we chose to follow the empirically validated methodology adopted by later methods as seen in \cite{liang2016bachbot}, \cite{sun2016composing}, \cite{colombo2016algorithmic} and \cite{yang2017midinet} where the network was able to learn low-level harmonic correlations between notes without the need for building them into the representation.\par

We represent a single pitch value by a one-hot encoded vector of length equal to the number of pitches playable on most pianos(88, from A0 to C8).

Varying approaches are taken for representing note durations. Some eschew the need for one by sampling the input tracks at a fixed rate. \cite{eck2002first} sample by {\em quavers}(eighth notes), while \cite{liang2016bachbot} samples by {\em demisemiquavers}(thirty-second notes). Some earlier monophonic approaches such as \cite{mozer1991connectionist} and \cite{franklin2004recurrent} used compressed binary vectors to encode note duration. \cite{sun2016composing} use a 30-bit vector to explicitly encode note lengths from a {\em semiquaver}(sixteenth note) to a {\em breve}(two whole notes). \par
We chose to use explicit duration encoding instead of the sampling method, firstly since it simplifies the multi-stream representation that we use, and secondly since forcing the network to learn patterns in note duration distributions was expected to help later during the {\em plan} interpolation process. Further, any errors in the sequential counting task resulting from the sampling approach can collapse the rhythm of a polyphonic composition. \par
We encode the durations as a one-hot vector, with each bit corresponding to a note duration that is frequently observed in the musical corpus that we use. 

\subsection{Existing Representations for Polyphonic Music}
A widely used representation for polyphonic music is the piano roll representation. Here, each bit in a binary vector represents the On/Off state of a pitch. The symbolic music is sampled at a fixed rate, and a {\em Sustain} pitch indicates whether or not the previous note is continued. \par
After this, the learning algorithm may either treat each note as a binary classification problem(\cite{boulanger2012modeling}), or sequentialize the notes(\cite{liang2016bachbot}) to produce $P$ $n_p$-class classification problems where $P$ is the instantaneous polyphony and $n_p$ is the number of possible pitches.  \par
\cite{chu2016song} and \cite{yang2017midinet} divide pop music into a melody track and a chord track, and use separate one-hot encodings for these tracks. However, this representation is limited to the class of songs that have a separate chord track that uses a standard set of multi-note chords for harmony. \par
\cite{hadjeres2017deepbach} use a pitch representation that has some similarities to our proposed representation. The major differences lie in our use of an explicit duration representation, the resolution of resulting incompatibilities and the imposition of pitch ordering.

\subsection{The Multi-Stream Note Representation for Polyphonic Music}
In the multi-stream representation which is a novelty arising from this work, we represent polyphonic music as a small set of streams, all of which are monophonic. Each stream has a pitch and duration component tied to it. This representation can embody polyphonic music of all forms and is limited in this process only by the number of streams $n_s$ and the set of permissible durations $\textbf{d}$.
\begin{itemize}
\setlength\itemsep{-0.1em}
\item During transcription, each incoming note in a polyphonic track is allotted to the lowest possible stream.
\item Multiple notes in a note-set are sorted in descending order of pitch to encourage the localization of closely spaced pitches in the same stream over time.
\item We introduce a {\em Rest} pitch value(111 here) to represent time-steps when a stream is not sounding any note.
\item We also use a {\em Sustain}\footnote{The occurrence of the Sustain pitch is deterministic and hence does not unnecessarily complicate the learning task.} pitch value(110 here) to handle cases where at the start of a note-set, a particular stream is expected to continuously play a previously started note rather than strike it again.
\item  Rest notes are filled into streams such that they do not have to be sustained in the next note-set.
\end{itemize}

\begin{figure}[h]
  \begin{center}
    \resizebox{80mm}{!} {\includegraphics *{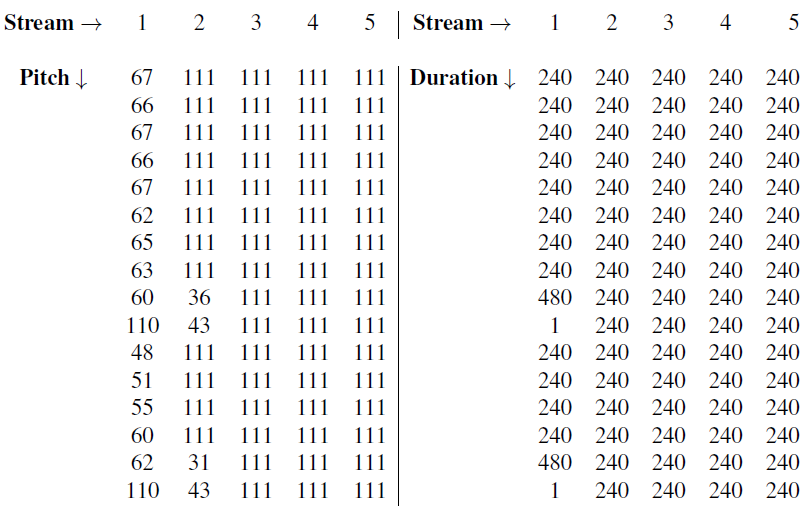}}
    \caption {\textbf{The first 16 note-sets from F\"ur Elise in the Multi-Stream Representation. Symbolic Pitches and Durations are depicted.}}
  \label{fig:multistreamElise}
  \end{center}
\end{figure}

\textbf{Fig \ref{fig:multistreamElise}} presents an example with the first few notes of Beethoven's F\"ur Elise in the multi-stream representation. This representation overcomes the following shortcomings of the piano roll representation and greatly reduces the sparsity of the input space:
\begin{itemize}
\setlength\itemsep{-0.1em}
\item The piano roll representation produces an extreme class imbalance for most pitches, since predicting each pitch is a binary classification problem, and since any given pitch is off during most samples. The lower streams in the multi-stream representation are spread very well across the various playable notes(classes), and any imbalance is by virtue of a particular note being played frequently.
\item The multi-stream representation produces a division of responsibilities among streams where the lowest stream generally learns to play the central melody, and higher notes learn the harmony or the ornamental notes.
\item Very few, if any single-instrument compositions will have more than 15-20 notes being played simultaneously. \cite{boulanger2012modeling} note that on a large collection of polyphonic music that included hundreds of musical pieces from different genres of music, the maximum polyphony was only $15$, with the average being $3.9$. The piano roll representation however, allows for the  unnecessary freedom of playing upto $n_p(88)$ simultaneous notes at the cost of increasing the sparsity of the space and the probability of incompatible pitches\footnote{A crude workaround for this is to impose an upper limit on the number of simultaneously sounded notes during the composition process}.
\end{itemize}

In the ordinary piano roll representation, if $T$, $S$, and $n_p$ are composition length, sampling rate, and the number of pitches respectively, then the number of possible compositions $N_1$ is given by
\begin{align*}
N_1 = 2^{\frac{T}{S}n_p}
\end{align*}
In the multi-stream representation with $n_s$ streams and with $\textbf{d} = {d_0, d_1, ... d_{n_d}}$ as the available set of note durations, a composition of time $T$ can be made up of several different combinations of durations.
Through numerical simulations, we observed that on reasonable choices of $\textbf{d}$ such as the {\em semiquaver} to {\em breve} set used in \cite{sun2016composing}, and our frequent durations set, the total number of possible compositions of length $T$ is greatly dominated by those that fully use the shortest durations for all notes\footnote{This is since the minimum duration is much smaller than the next available duration}. \par
The approximate number of compositions $N_2$ is then
\begin{align*}
N_2 \approx (n_p)^{n_s\frac{T}{d_0}}
\end{align*}
The logarithmic ratio, $R = log\frac{N_1}{N_2}$ for the representative values of $S = d_0 = \frac{1}{16}$, $n_s = 5$ and $n_p = 88$ is $268.29T$. In other words, the sparsity of the composition space in the piano roll representation increases $10^{268.29}$ times faster than the multi-stream representation for every whole note. We noted experimentally that during the composition phase, the neural network classified notes with much higher confidence with the multi-stream representation than with the piano roll representation.
\par

\subsection{Plan and Plan Interpolation}
The practice of using an additional set of variables called a {\em plan} to differentiate subsequences taken from different songs has previously been used in \cite{todd1989connectionist}, and later in \cite{franklin2001multi}. Both these approaches use a one-hot encoded binary vector to provide information on which song the given subsequence is from. During the composition phase, \cite{todd1989connectionist} uses interpolated plans to force creativity, while \cite{franklin2001multi} turns on all the bits in the plan. \cite{todd1989connectionist} notes that compositions inherits attributes from the songs that have high weights in the plan and offer the following paraphrased analysis of the plan inputs: The activations from the plan inputs act as biases on the hidden units, causing them to compute different functions of the context inputs. The context inputs can be thought of as points in a higher-dimensional space, and the hidden units act as planes to cut this space up into regions for different outputs. \par
This implies that switching a single bit in the plan input will cause significant alterations in the network's behaviour, and this is indeed corroborated by our experiments. \par
We first re-validated the attribute inheritance through experiment when we observed that major parts of the composition’s pitch and note duration distributions are combinations of strongly(with very high/low relative frequency) observed
attributes in the songs corresponding to the On bits. \par
As opposed to \cite{todd1989connectionist}, we limited the search space from $R^n$ to $\{(p_1, p_2, ... p_n) : p_i \in \{0,1\}\}$. This is since it transforms the compositional attribute inheritance from a parametrized setting to that of a competitive, dominance-based process. Here, multiple songs from the plan will compete or collaborate to influence the composition, as per their own mismatching or matching attributes. \par
The qualities and advantages of the {\em plan} space have been largely unexplored until now, especially its properties when enlarged by using a large number of songs. We present the first instance of a method that can intelligently explore interesting points in this space without undertaking an exhaustive search. We also show later with a few examples that the attribute inheritance goes beyond simple properties like pitch and duration distribution.

\section{LSTM Neural Network Training and Results}

\subsection{Training Data, Inputs and Targets}
We used MIDI versions\footnote{from http://www.piano-midi.de} of a selected set of solo piano compositions from Beethoven, Mozart, Lizst, Bach and Alb\'eniz. Since these songs all have varying tempos, we quantized the tempo values to ensure that a small, common set of note durations would represent note lengths in all these tracks. We applied further temporal quantization to adjust misaligned notes caused by human performance. \par
The input during each prediction task is a sequence of $l_c$ note-sets in the multi-stream representation, where $l_c$ is the context length for predicting the next note-set. Given these, the neural network is expected to predict the pitch and duration values for all $n_s$ streams in the next note-set. 

\subsection{Neural Network Structure and Training}
We observed that using shared LSTM layers for both the pitch and duration inputs resulted in faster training and better generalization than using separate layers for these inputs. Following the $n_l$ LSTM layers each with $n_u$ LSTM units, we added $2n_s$ fully connected layers, all using the softmax activation function to predict the pitch and duration values for each stream. The loss function $L$ was the sum of cross-entropies for each classification problem, i.e.
\begin{equation*}
    L = - \frac{1}{n_s}\sum\limits_{i=1}^{n_s} \sum\limits_{j=1}^{n_p} y_{p,i} \log \hat{y_{p,i}} - \frac{1}{n_s}\sum\limits_{i=1}^{n_s} \sum\limits_{j=1}^{n_d} y_{d,i} \log \hat{y_{d,i}}
\end{equation*}
We used the Adam Optimizer introduced in \cite{kingma2014adam} for gradient descent, and a batch size of 64. We used the Keras software library to implement the LSTM neural network. We perform a grid search over $l_c \in \{10,15,20,25,30\}$, $n_l \in \{1,2,3\}$, and $n_u \in \{100,200,300,400\}$ while minimizing the loss function defined above. We obtained $l_c = 20$, $n_l = 2$ and $n_u = 300$ as the optimal values.

\subsection{Network Learning over Epochs}

\begin{figure}[h]
  \begin{center}
    \resizebox{80mm}{!} {\includegraphics *{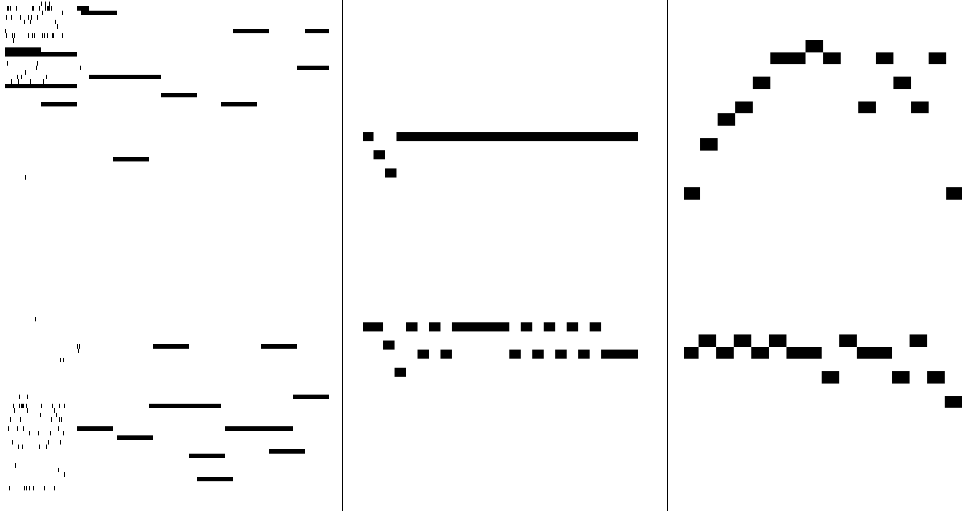}}
    \caption {\textbf{Compositions from the network at 4 Epochs(left), 20 Epochs(center) and 50 epochs(right). Y-Axes are not to equal scales.}}
  \label{fig:compositionVsEpoch}
  \end{center}
\end{figure}

\textbf{Fig \ref{fig:compositionVsEpoch}} shows the clear progress in the network's learning over training epochs. \par
At 4 epochs, the network has simply identified the pre-eminence of treble and bass parts in most of the training songs and emulates this. The composition is arrhythmic, with note starts and endings mismatched throughout the sample. \par
At 20 epochs, we observe significant improvements in the rhythm, and there are even symmetric movements between the treble and the bass components. There is also a rudimentary understanding of melodic intervals and tonality. However, there is little melodic or harmonic variation in the composition at this point. \par
At 50 epochs, the melodic and harmonic complexity greatly evidence themselves in the compositions. There are occasional contrapuntal sections(not seen here) where two separate melodies intertwine harmonically, and the rhythm is well-maintained. \par

\section{Composition}
We follow the sequential composition process used by \cite{todd1989connectionist}, \cite{eck2002first}, \cite{franklin2004recurrent} and \cite{sun2016composing} among others. We compose one note-set in every step, then append this note-set to the context for predicting the next note-sets. We seed the network with an initial sequence as is the practice in previous work. \par
For a given contextual sequence of past note-sets, the neural network outputs probability distribution vectors for the pitch and duration values for every stream in the next note-set: $\textbf{p}_{p,1}(x_{t+1}),\textbf{p}_{p,2}(x_{t+1}),... \textbf{p}_{p,n_s}(x_{t+1})$ and $\textbf{p}_{d,1}(x_{t+1}),\textbf{p}_{d,2}(x_{t+1}),... \textbf{p}_{d,n_s}(x_{t+1})$. \par
From these $2n_s$ distributions, we sample one value each, thus defining the entire note-set.

To exert control over the incidence of low-probability pitches and durations in the compositions, we modify the neural network's {\em softmax} output with the Boltzmann Distribution before sampling from it. A parameter called the {\em Temperature} is used to this end.
\begin{equation*}
    p_{i, new} = \frac{e^{\frac{\log p_i}{T}}}{\sum\limits_{j=1}^{n_p} e^{\frac{\log p_i}{T}}}
\end{equation*}
The probability distribution varies between a deterministic point for $T=0$, $\{p_i\}$ itself for $T=1$ and the discrete uniform distribution for $T=\infty$. \par
At low {\em temperatures}, the network often makes predictions that are "safe" and have low variation. With an increase in {\em temperature}, the predictions become varied, at the risk of selecting notes that are not apt for the composition. \par
We use this {\em temperature} during the reinforcement learning phase as it allows us to tune melodic, harmonic and rhythmic variation with our need to reduce poorly placed notes.

\section{Applying Reinforcement Learning}

\subsection{Applicability of Reinforcement Learning}

A neural network trained to reproduce existing musical pieces when given completely unseen inputs will just output compositions that have high probabilities according to the distribution it previously learnt. While they do tend to be guided by the conditional probabilities observed in existing music, their original compositions may not possess positive attributes that we expect from them. The compositions may contain unwanted quirks that simply maximize the joint probability according to the network, for instance, a small cycle of somewhat pleasing notes that repeats indefinitely. \cite{sun2016composing} and \cite{jaques122016tuning} note that the music generated through a multi-step process from RNNs lack global structure and are overly repetitive. \par
These observations establish the strong need for having feedback and conditioning on the neural network's compositions. \cite{briot2017music} observe that "the reinforcement strategy allows to combine arbitrary user given control with a style learnt by the recurrent network". \par

\subsection{Rewards and Penalties}
We define the following rewards and penalties (denoted (\textbf{+}) and (\textbf{-}) respectively), basing upon rules used previously in \cite{franklin2001multi}, \cite{sun2016composing} and \cite{jaques122016tuning}, and tailoring it to our specific requirements:
\begin{enumerate}[itemsep=0.0mm]
\item (\textbf{+}) Occurrence of dyads, triads and seventh chords.
\item (\textbf{+}) High pitch entropy.
\item (\textbf{-}) Overuse of very short or long note durations.
\item (\textbf{-}) Multiple identical note-sets in sequence.
\item (\textbf{-}) Rests form too large a part of the played note-sets.
\item (\textbf{-}) Large peaks in the normalized cross-correlations\footnote{Cross-correlation is performed with the pitch values represented in one-hot encoding. Only note-sets with overlapping pitches must contribute to the cross-correlation.} between pitch values in subsequences from the composition and songs in the training set\footnote{This penalty was added to suppress compositions that plagiarize parts of the training data.}.
\end{enumerate}

The numerical parameters and thresholds for these rewards and penalties were chosen by maximizing the aggregate reward on a set of compositions that were manually determined to be pleasant.  

\subsection{Problem Modelling}

In a novel exercise, instead of directly altering the weights of the previously trained polyphonic reproduction network or cascading note-level sampling networks that learn through RL, we take a different approach. We use the powerful, varied, high-level control offered by the {\em plan} input and the {\em temperature} parameters to surface combinations that result in compositions that conform to our expectations.\par
In this case, the problem of searching through the space of {\em plans}  and {\em temperatures} can be restated as a Markov Decision Process(\cite{puterman2014markov}):
\begin{itemize}
\setlength\itemsep{-0.1em}
\item {\em Plan} and {\em temperature} inputs together form a state $s$.
\item At every state, the allowed actions\footnote{In a simplification, the state transitions upon taking actions are all deterministic.} A involve either switching a particular bit in the {\em plan}, or altering the {\em temperatures}.
\item The reward is obtained by composing a song with the selected {\em plan} and {\em temperatures} and evaluating it.
\end{itemize}

We use the mathematical framework and the iterative algorithm described in \cite{watkins1989learning} and \cite{watkins1992q} to learn the Q-Values. 
We use a three-layer neural network as a function approximator(\cite{crites1996improving}) to store and predict the Q-Values. This methodology has sometimes been observed to improve learning by using previous experience towards predicting the utility of unseen states.

\subsection{Implementation}
We implemented the Q-Learning algorithm in Python and used the Keras software library to implement the function approximator neural network. We used Mean Squared Error(MSE) and Stochastic Gradient Descent(SGD) to train the neural network. We also performed a grid search to find the optimal hyperparameters among discount factor $\gamma \in \{0.2, 0.4, ... 1.0\}$, learning rate $\eta \in \{10^{-3}, 2.5\times10^{-3}, 5\times 10^{-3}, 7.5\times 10^{-2}, 10^{-2}\}$ and number of hidden neurons $n_h \in \{10,20,30,40\}$. We obtained $\gamma = 0.8$, $\eta = 0.075$ and $n_h = 20$ as the optimal values.

\subsection{Comparison of Average Compositional Quality with and without RL}

\begin{figure}[h]
  \begin{center}
    \resizebox{80mm}{!} {\includegraphics *{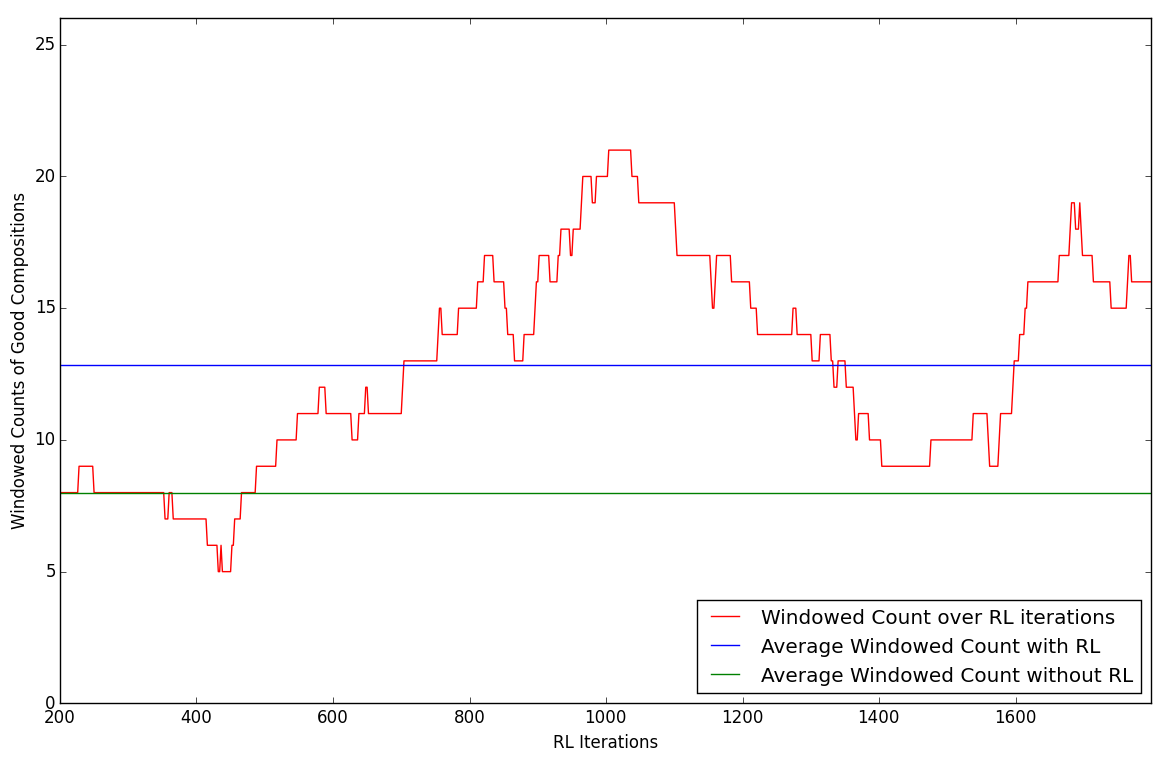}}
    \caption {\textbf{Moving Windowed Count of Good Compositions against RL Iterations}}
  \label{fig:goodCountVsTime}
  \end{center}
\end{figure}

The plot in \textbf{Fig \ref{fig:goodCountVsTime}} illustrates the progress and utility of the learning process. It shows the count of good compositions\footnote{A good composition is taken as that which fetches a reward greater than $5.0$ out of a maximum of $7.0$.} in a moving window 200 iterations wide.  From the strong upward movement of the average count line, we see clearly the advantage offered by reinforcement learning over a random {\em plan} selection approach. \par
\newcolumntype{R}[1]{>{\raggedright\let\newline\\\arraybackslash\hspace{0pt}}m{#1}}
\begin{table}[h]
\begin{tabular}{R{3cm} c c}
\toprule
\multicolumn{1}{c}{} & \multicolumn{2}{c}{\textbf{Average over 200 Compositions}}\\
\cmidrule{2-3}
\textbf{Attribute} & \textit{With RL} & \textit{Without RL} \\ \midrule
Dyads, Triads and Seventh Chords & \textbf{0.24} & 0.12 \\
Pitch Entropy & \textbf{0.82} & 0.71 \\
Very short/long duration incidence & 0.05 & 0.08 \\
Repeated Identical Note-sets & \textbf{0.06} & 0.20 \\
Aggregated Rest Duration & \textbf{0.08} & 0.14 \\
Rest Count & 0.07 & 0.11 \\
Cross-Correlation peak & \textbf{0.13} & 0.2 \\
\bottomrule
\end{tabular}
\caption{\textbf{Improvements in compositional attributes when using Reinforcement Learning}}
\label{table:rlImprovementsTable}
\end{table}
Detailed per-attribute improvements are shown in \textbf{Table \ref{table:rlImprovementsTable}}. The averages in the table are on absolute values for the entropy and the cross-correlation peak, and relative values for the other attributes.

\section{Qualitative Analysis of the Compositions}

\begin{figure}[h]
  \begin{center}
    \resizebox{55mm}{!} {\includegraphics *{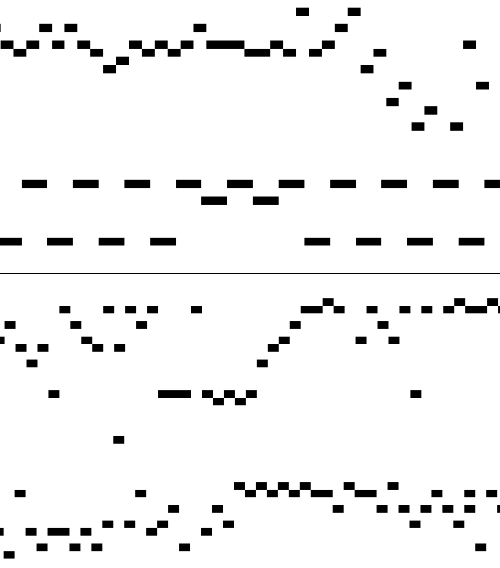}}
    \caption {\textbf{Two example Compositions dominated by F\"ur Elise(top) and Espana Op. 165 Prelude(bottom)}}
  \label{fig:exampleCompositions}
  \end{center}
\end{figure}

Two interesting subsequences from \textbf{Amadeus}' compositions are shown in \textbf{Fig \ref{fig:exampleCompositions}}. The sequence on the top was composed with the {\em plan} bits corresponding to F\"ur Elise and Bach's BWV 850 set to 1.0. It shows a clear melodic sequence with an accompanying harmony, quite reminiscent of F\"ur Elise. The pitch distribution, however, is strongly influenced by BWV 850, with most notes drawn from the C Major scale. It is also interesting to note that both the melody and the harmony follow and maintain distinct note lengths that together maintain the rhythm. \par
The lower piece was composed with the {\em plan} bits corresponding to the $3^{rd}$ Movement of Appassionata and the Prelude from Alb\'eniz's Espa\~na set to 1.0. The composition inherits the contrapuntal qualities of both these songs. There are sections in the composition where the melody frequently uses intervals from Espa\~na. Once again, there is clear differentiation between the contrapuntal melodies, and the rhythm is maintained almost perfectly. \par
However, higher-level musical qualities such as organized repetition with variations, surprise, tension/resolution, and well-designed climactic sequences are still out of reach at present. Identifying techniques to embody these will be the focus of our future research.

\section{Conclusion, Audio Tracks and Additional Material}
We presented a novel multi-stream representation that takes advantage of polyphonic music's structure to simplify the learning problem that we are solving here. We used this representation to train an LSTM Neural Network, then in a unique approach, applied Reinforcement Learning to select high-level network configurations(rather than directly modify the network weights or its outputs) that maximize a set of attributes that we expect from pleasant algorithmic music. We also demonstrated the significant utility of the previously introduced {\em plan} space towards producing controlled, but diverse compositions. \par
Our system demonstrates its ability to emulate some of the notable features of music composed by human composers such as harmony, melodic complexity, tonality, counterpoint, rhythm, and the proper use of treble and bass components. While some of these properties have been displayed by previous research work in this field, the compositions produced by \textbf{Amadeus} present all of these as a coherent whole, thus producing music that is one more step closer to human-level composition. \par
In the future, we hope to first add more localized rewards and introduce an attack velocity for the played notes. We also hope to undertake a comparative rating test of our compositions on human audience(e.g. \cite{liang2016bachbot}, \cite{yang2017midinet} and \cite{hadjeres2017deepbach}). \par
Audio tracks of compositions, and additional material such as structural diagrams, datasets, etc. can be found at \url{https://goo.gl/ogVMSq}

\bibliographystyle{named}
\bibliography{ijcai18}

\end{document}